\RequirePackage{fix-cm}
\documentclass[smallextended]{svjour3}       
\smartqed  
\usepackage{graphicx}
\usepackage{url}
\usepackage[authoryear,longnamesfirst]{natbib}
\usepackage{wrapfig}
\usepackage{color}
\usepackage[linesnumbered,ruled,vlined]{algorithm2e}

\SetCommentSty{mycommfont}
\usepackage{algpseudocode} 
\usepackage{multirow}
\graphicspath{ {Figures/} }
 \journalname{Scientometrics}
\begin{document}

\title{Emerging trends and collaboration patterns unveil the scientific production in blockchain technology: A bibliometric and network analysis from 2014-2020}


\titlerunning{Blockchain bibliometrics}        

\author{Kiran Sharma \and Parul Khurana}

\authorrunning{K.Sharma et~al.} 

\institute{Kiran Sharma \at
               School of Engineering and Technology,  BML Munjal University, Gurugram, Haryana-122413, India \\
              \email {kiran.sharma@bmu.edu.in}
\and
Parul Khurana \at
              School of Computer Applications, Lovely Professional University, Phagwara, Punjab-144401, India \\
               \email{parul.khurana@lpu.co.in}              
}


\maketitle

\abstract{Significant attention in the financial industry has paved the way for blockchain technology to spread across other industries, resulting in a plethora of literature on the subject. This study approaches the subject through bibliometrics and network analysis of 6790 records extracted from the Web of Science from 2014-2020 based on blockchain. This study asserts (i) the impact of open access publication on the growth and visibility of literature, (ii) the collaboration patterns and impact of team size on collaboration, (iii) the ranking of countries based on their national and international collaboration, and (iv) the major themes in the literature through thematic analysis. Based on the significant momentum gained by the blockchain, the trend of open access publications has increased 1.5 times than no open access in 2020. This analysis articulates the numerous potentials of blockchain literature and its adoption by various countries and their authors. China and the USA are the top leaders in the field and applied blockchain more with smart contracts, supply chain, and internet of things.
Also, results show that blockchain has attracted the attention of less than 1\% of authors who have contributed to multiple works on the blockchain and authors also preferred to work in teams smaller in size.}

\keywords{Blockchian, Bibliometric analysis, Collaboration network, Country ranking, Thematic analysis, Open access}

\maketitle

\section{Introduction}	
\label{IN}

Blockchain~\citep{kube2018daniel} deployment since its initial perception was largely experimental. Various authors have strongly introduced the promising non-finance applications which have raised the special attention towards blockchain across the globe~\citep{white2017future}. Blockchain has been seen as a disruptive combustion engine with the potential to transform organizations in the current digital economy~\citep{prybila2020runtime} with features like validation of transactions, safeguarding of entries~\citep{olnes2017blockchain}, preservation of records, immutability, decentralization, consensus, and faster settlement. The idea of Satoshi Nakamoto (as an individual or a group) to send online payments directly from one party to another~\citep{nakamoto2008bitcoin}, without any interference from financial institutions, started the hype of blockchain in 2008. His solution described the first realization of the concept as a Bitcoin~\citep{bohme2015bitcoin},  which was intrinsically tied with blockchain technology~\citep{crosby2016blockchain}.

Since then the authors have demonstrated great interest in the blockchain literature based on features, principles, applications, adoption challenges, and opportunities to influence the potential of blockchain technology~\citep{yli2016current}. Various literature hoping for new industry driven solutions and promising significant business benefits~\citep{lopez201860} from blockchain technology have also emerged. Such literature have also highlighted innovations, emerging trends and corporate confidence integrated into the adoption of blockchain technology~\citep{hoffman2019scholarly}. Different authors have also analyzed blockchain as a consortium producing a greater momentum and specific problem solver for emerging economies in various countries~\citep{zeng2018bibliometric, firdaus2019rise}.

Despite various literature available on blockchain~\citep{dabbagh2019evolution, yalcin2021mining, guo2021bibliometric, tandon2021blockchain},  this study figured out the gaps in the existing study and presented the bibliometric and network analysis of literature to show the trends in open access publications and thematic analysis~\citep{velez2020thematical} of technology across countries. The impact of collaboration strength~\citep{sharma2021growth} and team size is analyzed from the author's collaboration network. Further, the share of national and international collaborations~\citep{lee2005impact} is analyzed from country collaboration networks and a country ranking is being built based on the national and international collaboration. The objectives of our study are outlined as:
\begin{enumerate}
\item Bibliometric analysis of the publications to present a comprehensive overview of the blockchain literature from 2014-2020.
\item To investigate the impact of open access publication on the growth and visibility of blockchain.
\item To investigate the impact of author's collaboration and team size from author's collaboration network.
\item To identify top countries based on their national and international collaboration from country collaboration network.
\item To build a country ranking based on their national and international collaboration.
\item To perform thematic analysis to demonstrate the usage of major themes in top countries.
\end{enumerate}

Further, the study is organized as follows: Section~\ref{sec1} is on data description and feature extraction.  Section~\ref{sec2} presents the view on the growth and impact of blockchain, highlighting the temporal distribution of publications along with publication categories and citations analysis.  Section~\ref{sec3} presents the author`s collaboration network analysis highlighting the author's contribution, strength,  and team size.  Section~\ref{sec4} is on the country collaboration network analysis highlighting the country-wise authors and leader, their national and international collaboration,  countries ranking based on their national and international collaboration, and thematic analysis of technology across countries.  Finally, Section~\ref{sec5} concludes the study.

\section{Methodology}
\label{sec1}
The methodology section covers two main stages (i) data collection,  and (ii) feature extraction to perform a structured review. 

\subsection{Data collection}  
Data is collected from Web of Science powered by Clarivate Analytics (\url{https://webofknowledge.com/}). The choice of the database is arbitrary.  This study preferred the WoS database because its authenticity and reliability ~\citep{gorraiz2016availability, martin2019google}.
The process started with the search of the keyword ``Blockchain'' in the whole WoS database as (KP=(blockchain) OR AK =(blockchain)). The query returned the 6790 records from 2014-2020. The data is extracted on Feb 2021.  The metadata contains the information of publication year, authors, affiliation, journal,  publisher, citations, downloads, research category, open access category,  publication language,  reprint address, etc.  The metadata is prepared on articles, reviews, letters, editorial materials,  and proceedings.
\subsection{Feature extraction}
Before starting with the bibliometric analysis,  we performed feature extraction on the data as follows:
\begin{itemize}
\item To find the trend of publications, all 6790 documents were arranged as per the year of publication.

\item To identify the open access designations of each paper,  the papers with Gold, Bronze and Green categories are filtered. This process returned 1597 (23.5\%) as Gold,  237 (3.5\%) as Bronze,  and 193 (2.8\%) as Green publications.  4763 (70.2\%) publications belong to the no open access category (NOA). 

\item To find the publications distribution as per the language of publication,  a publication count of each language is prepared.
\item To measure the impact of publications, the citations received and usage count of each article is observed.

\item The number of publications corresponding to authors and country was not available directly,  hence the information is extracted from the author's affiliation.  A total of 17686 unique authors and 106 countries were extracted from the data. Similarly,  the national and international contribution is also filtered from the affiliations.

\item To extract country-wise total publications and authors, we gave equal credit to multi- authored and affiliated publications. Hence, we came up with 9550 publications.

\item Countries information of corresponding authors is extracted from the reprint address. Also. country collaboration information is collected from the publication address.

\item Themes corresponding to the top 10 countries are extracted from every individual paper manually.
\end{itemize}

\section{Growth and impact of blockchain}
\label{sec2}
The results are presented in three phases. The first phase is on the growth and impact of blockchain which is discussed through the bibliometric analysis. This phase mainly highlights (a) the temporal distribution of the number of publications along with the categorization of the open-access documents; (b) the dominant publication language; (c) the documents category along with the number of publications in that category and citations received; and (d) the impact of the publications.

\subsection{Temporal distribution of publications}
\begin{figure}[h!]
\centerline{\includegraphics[width=0.8\linewidth]{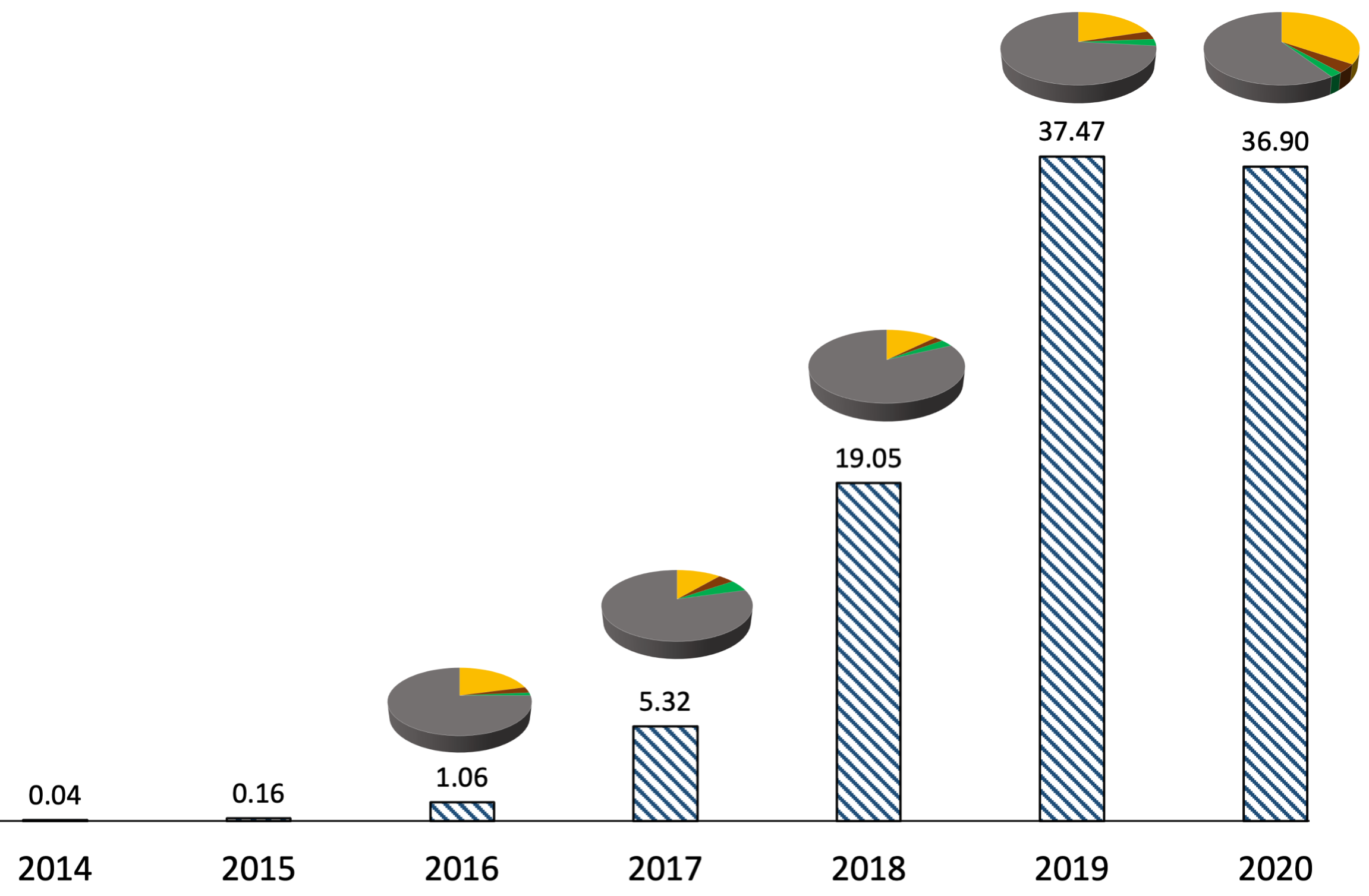}}
\caption{The bar plot represents the temporal trend of the number of papers published [\%] during 2014-2020.  The pie plot on each bar represents the distribution of the open-access publications.  Three categories of open access publications are Gold (colored in yellow),  Bronze (colored in brown), and Green (colored in green). The fourth category belongs to No Open Access (NOA) (colored in grey).}
\label{fig1}
\end{figure}

Literature growth of the blockchain has been studied by many researchers in the past~\citep{miau2018bibliometrics}; however, the relation of such growth with open access has not been analyzed earlier. Here, we analyzed the publication trend and its association with the open access publication.
Fig.~\ref{fig1} shows the number of papers published from 2014-2020 [\%].  In 2016,  the growth measured as 4.26\% as compared to the past performances, and it has gone up by three times in 2017 (13.73\%).  Hence, a rise in the number of publications can be seen from 2017 onward.  A sharp rise of publication count of 18.42\% can be seen in 2019; however,  a fall of 0.57\% is observed in 2020. The temporal trend also shows that studies on the blockchain have gained momentum in 2018.

There are numerous advantages of publishing open access~\citep{antelman2004open, martin2018evidence}.  In Fig.~\ref{fig1} the pies on top of each bar represent the distribution of publications in terms of the open access (OA) category.  There are three main open access categories- Gold (colored in gold), Bronze (colored in brown), and Green (colored in green),  and fourth belongs to no open access (NOA colored in grey). The open-access publications lead to the visibility of the research work and the acceptance towards this can be seen here. 

The contribution of Gold OA publications has risen as 20.8\% in 2016; however, in 2017, the share was 11.3\%.  In 2020 (34.6\%) this contribution hyped 1.5 times as compared to 2019 (20\%).  We also observed that with the rise in Gold OA publications, the NOA submission has gone down by 1.2 times in 2020 (59.3\%) as compared to 2019 (73.2\%).
Overall,  the average number of publications by Gold OA is 23.5\% with average citations of 8.4,  Bronze OA is 3.5\% with average citations of 6.6,  and Green OA is 2.8\% with average citations of 14.1. On the other hand,  NOA has a higher number of publications (70.2\%) with average citations of 6.4 (see Table.\ref{table1}).

\subsection{Measure publication impact through citations, usage count and open access}
\begin{figure}[h!]
\centerline{\includegraphics[width=0.55\linewidth]{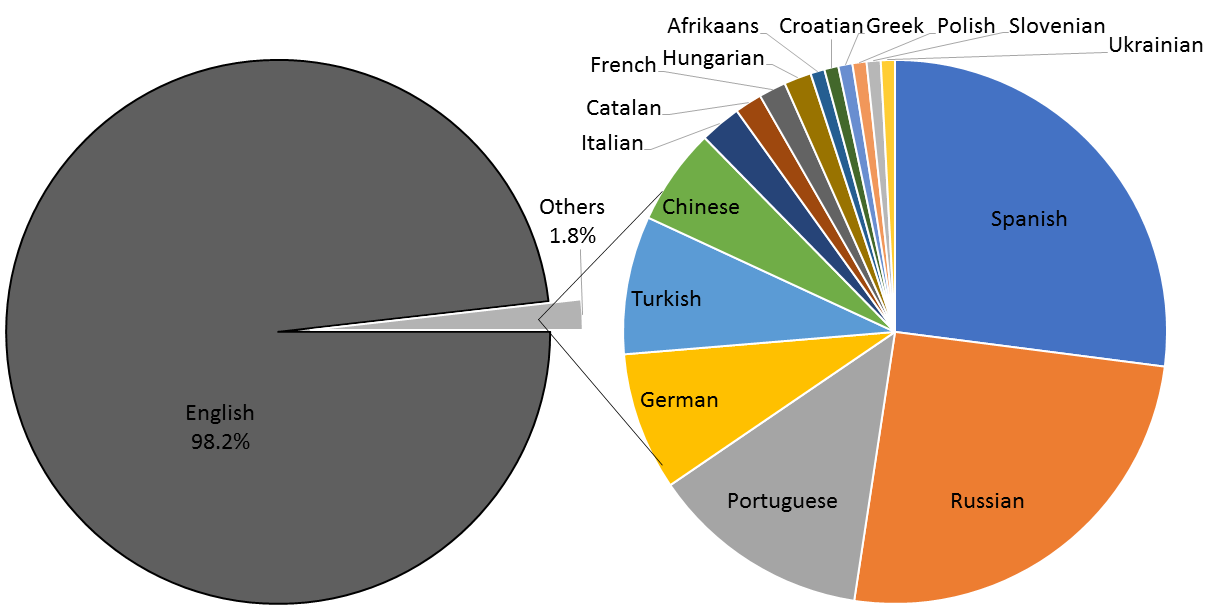}}
\caption{Pie plot shows the language-wise distribution. Papers got published in 17 languages; however, the major publication is in the English language followed by Spanish and Russian. }
\label{fig2}
\end{figure}

The quality and visibility of English publications are far better than any other language.  Further,  publications in English do not require any translation at the international level.  Hence,  even the countries where English is not the first language, authors do prefer to publish in English. This gives an opportunity to authors to present their work at international platform \citep{meneghini2007there}.  Papers on Blockchain are published in 17 languages but 98.2\% of them are published in the English language as shown in Fig.~\ref{fig2}. There are 34 papers published in Spanish, 32 in Russian, 17 in Portuguese,  and 11 in Turkish.  In total 1.8\% (122) papers are published in languages other than English and the majority are the articles.

\begin{figure}[h!]
\centerline{\includegraphics[width=0.95\linewidth]{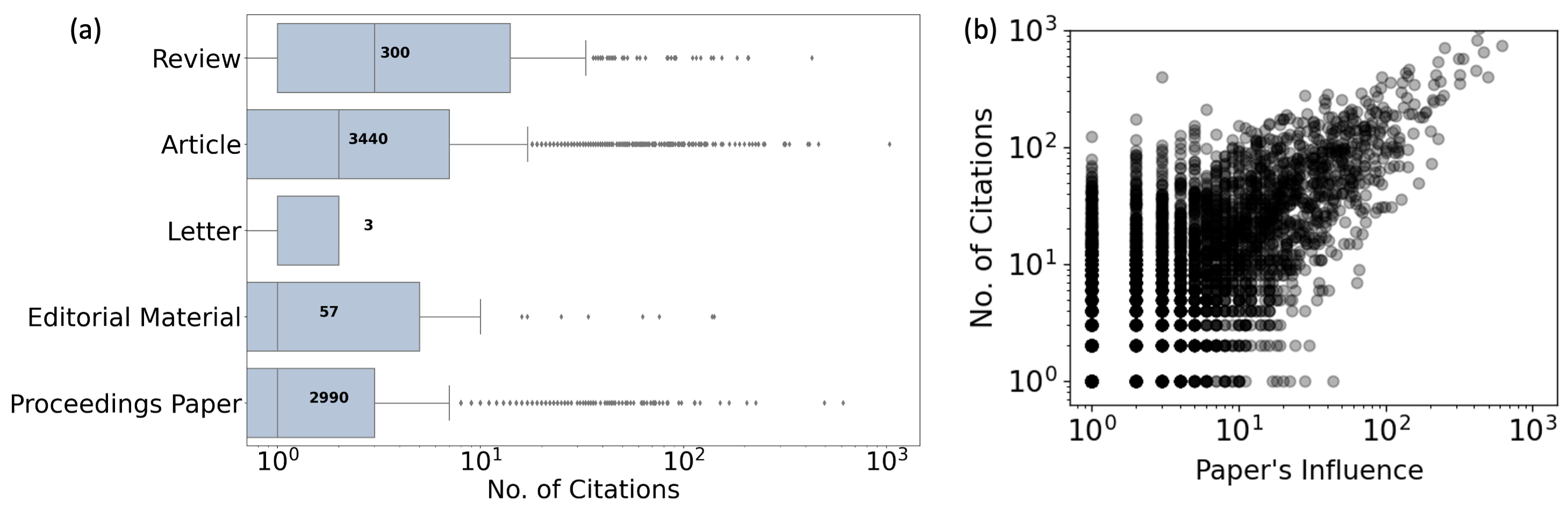}}
\caption{(a) The median number of citations received by different documents published during 2014-2020. The numeric value inside the box is the total number of published papers in that document category. (b) The number of citations versus the number of downloads.}
\label{fig3}
\end{figure}

A box plot of documents category in Fig.\ref{fig3} (a) shows the number of publications and corresponding citations.  49.7\% of publications have been submitted as articles,  44\% as conference proceedings, and 4.4\% as reviews.  The average number of citations received by articles is 9.1,  proceedings as 3.9  and reviews as 16.9.  Although the contribution as proceedings is nearly equivalent to articles, however, the average citations of proceeding articles is very low as compared to articles.  On the other hand, the review has the lowest count but the highest average citations among the three.  So, this raises a question that is the publication category of proceedings is NOA? So we have analyzed that only 7.7\% proceedings are published as OA. Hence, the visibility is low and so are the citations.  On the other hand, 47\% of articles are published as OA and 54\% reviews as OA which means the visibility is higher so are the citations.  Hence,  the impact of this distribution is clearly reflected in the citations.

The visibility, influence and the impact of the publications can be directly measured through the citations,  usage count, and open access~\citep{wang2016usage, lariviere2018authors}.
Fig.\ref{fig3} (b) shows the relationship between the number of citations and the paper's influence.  The paper's influence is measure through the usage count provided by WoS.  It provides insight into how much interest a publication has generated among users of the platform. The visibility of the publication is twice the actual citations, i.e.,  the average number of citations received by blockchain-based publications is 7.1 and the average number of downloads received by blockchain-based publications is 16.3.  Overall only 2.9\% of publications are highly cited that received more than 50 citations which are mainly as articles and reviews.  Mainly,  the highly cited papers have been viewed 2.5 times than it received citations. 

During 2017-18, OA count has decreased but during 2019-20 it has gained the momentum (see Fig.~\ref{fig1}). In 2019, the share of OA publications was 26.8\% and 40.7\% in 2020. Base on this,  in 2016 the average number of citations received by NOA publications is 54.7 and OA is 45.3.  As we can see that 25\% of OA publications have on an average 45.3 citations which itself reveals the impact of OA.  Similarly in 2019, 26.8\% of OA publications received on an average 41.6 citations.
Further, we analyzed that 43.7\% of publications are published by IEEE, followed by 9.9\% by Springer, 9.7\% by Elsevier.  The broader categorization of disciplines highlights that the major contribution of the blockchain is in the field of \textit{Engineering \& Technology} followed by the second highest in \textit{Social Sciences} (see Table.~\ref{table1}).

\begin{table}[h!]
\caption{Statistics}
\begin{tabular}{ccl|l|c|l|l|c|}
\cline{1-2} \cline{4-5} \cline{7-8}
\multicolumn{1}{|c|}{\begin{tabular}[c]{@{}c@{}}Conference\\ Proceedings {[}\%{]}\end{tabular}}    & \multicolumn{1}{c|}{44}   &  & \multicolumn{2}{c|}{\textbf{\begin{tabular}[c]{@{}c@{}}Discipline-wise\\ Publications {[}\%{]}\end{tabular}}} &  & \multicolumn{2}{c|}{\textbf{\begin{tabular}[c]{@{}c@{}}Top 5 Publishers\\ {[}No. of Papers in \%{]}\end{tabular}}} \\ \cline{1-2} \cline{4-5} \cline{7-8} 
\multicolumn{1}{|c|}{\begin{tabular}[c]{@{}c@{}}Publications with\\ Funding {[}\%{]}\end{tabular}} & \multicolumn{1}{c|}{49.5} &  & \begin{tabular}[c]{@{}l@{}}Engineering \&\\ Technology\end{tabular}                   & 75.4                  &  & IEEE                                                       & 43.7                                                  \\ \cline{1-2} \cline{4-5} \cline{7-8} 
\multicolumn{2}{|c|}{\textbf{\begin{tabular}[c]{@{}c@{}}Highly Cited\\ Publications {[}\%{]}\end{tabular}}}                    &  & Social Sciences                                                                       & 15.5                  &  & Springer                                                   & 9.9                                                   \\ \cline{1-2} \cline{4-5} \cline{7-8} 
\multicolumn{1}{|c|}{\begin{tabular}[c]{@{}c@{}}Below 50\\ Citations\end{tabular}}                 & \multicolumn{1}{c|}{97.1} &  & Natural Sciences                                                                      & 6.2                   &  & Elsevier                                                   & 9.7                                                   \\ \cline{1-2} \cline{4-5} \cline{7-8} 
\multicolumn{1}{|c|}{\begin{tabular}[c]{@{}c@{}}Above 50\\ Citations\end{tabular}}                 & \multicolumn{1}{c|}{2.9}  &  & \begin{tabular}[c]{@{}l@{}}Medical \&\\ Health Sciences\end{tabular}                  & 1.9                   &  & MDPI                                                       & 7.3                                                   \\ \cline{1-2} \cline{4-5} \cline{7-8} 
\multicolumn{1}{l}{}                                                                               & \multicolumn{1}{l}{}      &  & Agricultural Sciences                                                                 & 0.5                   &  & ACM                                                        & 5.1                                                   \\ \cline{4-5} \cline{7-8} 
\end{tabular}
\label{table1}
\end{table}


\section{Author's collaboration network analysis}
\label{sec3}
The contribution of the individual author can be analyzed through the author's collaboration network~\citep{sharma2021growth}.  Here,  the author's contribution analysis phase highlights (a) the author's collaboration network; (b) statistical analysis of collaboration network; and (c) authors contribution in several publications,  author pair and the publications,  and the number of publications in teams.

\subsection{Author collaboration network}

\begin{figure}[h!]
\centerline{\includegraphics[width=0.95\linewidth]{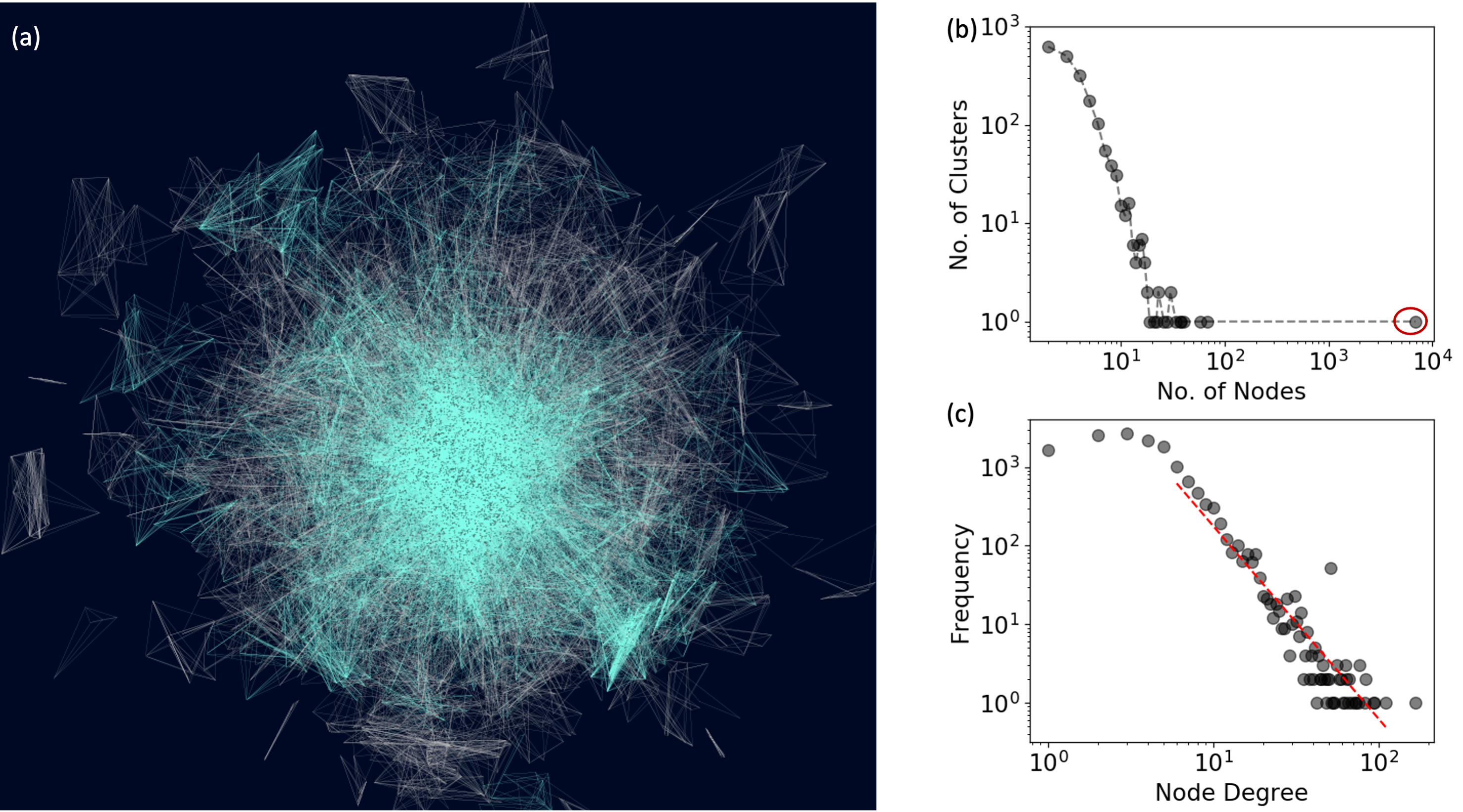}}
\caption{(a) An undirected author's collaboration network consists of 14748 nodes and 45919 edges. The nodes in the network are authors and edges are the collaboration link among two authors.  The nodes in the network are colored in grey and light sky blue. The nodes colored in light sky blue corresponds to a giant component of the network and the grey nodes are other weakly connected components.  The network is constructed with open-source software \textit{Gephi}. (b) The distribution of 1947 weakly connected components with the giant component marked in the red circle. The giant component consists of 46.3\% of network nodes.  (c) The degree distribution of nodes obeys the power law with exponent $-2.45$  and $R^2 = 0.88$. }
\label{fig4}
\end{figure}

An undirected network of the author's collaboration is shown in Fig.\ref{fig4} (a).  Out from 17686 authors, only 14748 authors shared at least one collaboration.  So, the network consists of 14748 nodes and 45919 edges. The nodes in the network represent the authors and the edges represent the possible collaborations between two authors.  A link appears between two nodes when two authors have a joint publication. The size of the node reflects the number of connections an author has with others (degree) and the width of the edge reflects the number of times the same two authors collaborate (strength). The overall network comprises 1947 weakly connected components (subgraphs colored in light grey) and out of these the giant component (the biggest connected subgraph) consists of 46.3\% of nodes (colored in light sky blue). The number of nodes in the giant component is highlighted in a red circle in Fig.\ref{fig4} (b). The distribution shows that one component/subgraph has nearly $10^4$ nodes and a couple of components are in the range of $10^2$ and the rest are below that.  The degree of the network represents the number of connections an individual node has in the network as shown in Fig.\ref{fig4} (c). The degree distribution follows the power law with exponent $-2.45$  and $R^2 = 0.88$. 10\% of nodes have degrees greater than 10 and a couple of nodes have degrees higher than $10^2$.

\begin{table}[h!]
\caption{Author's collaboration network statistics.}
\begin{tabular}{|l|c|l|c|c|}
\hline
\multicolumn{5}{|c|}{\textbf{Network Statistics}}                                                                             \\ \hline
Nodes                       & 14748                     &  & \multicolumn{1}{l|}{No. of Clusters}                     & 2226  \\ \cline{1-2} \cline{4-5} 
Edges                       & 45919                     &  & \multicolumn{1}{l|}{No. of Weakly Connected Components}  & 1947  \\ \cline{1-2} \cline{4-5} 
Avg. Degree                 & 5.2                       &  & \multicolumn{2}{c|}{\textbf{Team Size {[}No. of Papers in\%{]}}} \\ \cline{1-2} \cline{4-5} 
Avg. Weighted Degree        & 6.2                       &  & Single Author                                            & 38.7  \\ \cline{1-2} \cline{4-5} 
Network Diameter            & 16                        &  & Two Authors                                              & 28.9  \\ \cline{1-2} \cline{4-5} 
Avg. Path Length            & 5.7                       &  & Three Authors                                            & 16.5  \\ \cline{1-2} \cline{4-5} 
Avg. Clustering Coefficient & \multicolumn{1}{c|}{0.8} &  & More Than Five Authors                                  & 3.5   \\ \cline{1-2} \cline{4-5} 
\end{tabular}
\label{table2}
\end{table}
\subsection{Analysis of author's publications and collaboration patterns}
\begin{figure}[h!]
\centerline{\includegraphics[width=0.95\linewidth]{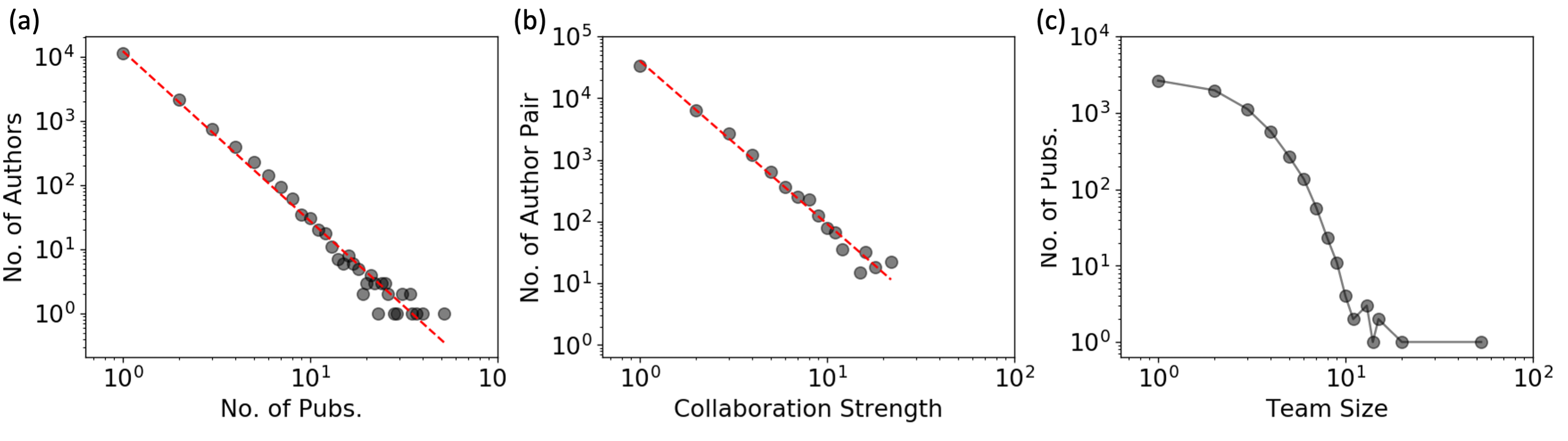}}
\caption{(a) Number of publications per author follows power law with exponent $-2.65$ and $R^2 = 0.99$.  (b) The author pair and the collaboration strength follows the power law with exponent $-2.64$  and $R^2 = 0.99$.  (c) Team size corresponding to the number of publications.}
\label{fig5}
\end{figure}

In the interdisciplinary era, the increasing trend of research collaboration has increased the productivity of the science~\citep{lee2005impact}.
The contribution of all authors has been measured in terms of the number of publications and collaboration is presented in Fig.~\ref{fig5} (a). The distribution of author's contribution follows the power law with exponent $-2.65$ and $R^2 = 0.99$.  The analysis reflects that 73.7\% authors have published a single paper on blockchain till now, and 14\% published two papers whereas more than 10 papers have been published by $<1\%$ of authors.  Also, the most active collaboration is being highlighted in Fig.~\ref{fig5} (b). Here, the relationship between the author pair and the collaboration strength follows the power law with exponent $-2.64$  and $R^2 = 0.99$.  The analysis shows that 73.5\% of author pair has published a single paper, 14\% of author pair has published two papers,  5.7\% published three,  and 0.5\% has published more than 10 papers together.  This highlights that only a few authors are frequently doing the research on blockchain.
Further, the team size contribution reflects the collaborative patterns in Fig.~\ref{fig5} (c).  38.7\% papers are written by the individual author, 28.9\% are written by the team of two authors,  16.5\% are written by the team of three authors,  and 3.5\% papers are written by the team of more than 5 authors.

\section{Country collaboration and thematic analysis}
\label{sec4}
The progress of an individual country in terms of publications, collaboration,  etc.  can be beautifully analyzed through the country's collaboration network.  Here, the country contribution analysis phase highlights (a) the country collaboration network; (b) country-wise the number of authors and publications; (c) ranking of countries based on national and international collaboration; and (d) thematic analysis of technology across countries.

\subsection{Country collaboration network}
\label{country}
\begin{figure}[h!]
\centerline{\includegraphics[width=0.9\linewidth]{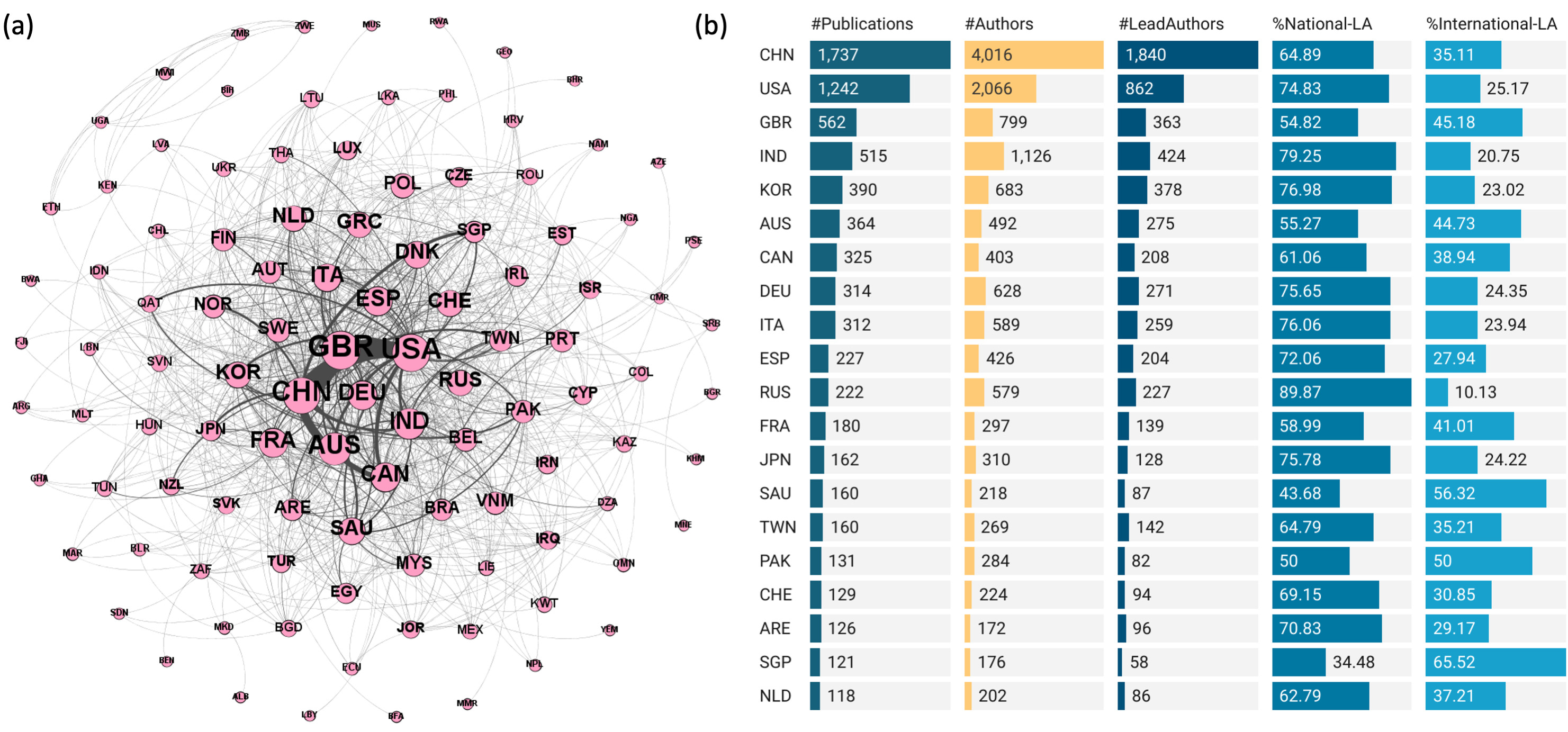}}
\caption{A demonstration of the country collaboration network with 106 nodes and 4009 edges. The nodes in the undirected weighted network represent the country and edges represent the collaboration.  The size of the nodes is dependent on the node degree and the width of the edges represents the strength of collaboration.  The highest collaboration is shared by the USA and UK followed by China, India, Australia, etc. The network is constructed with open source software \textit{Gephi}.}
\label{fig6}
\end{figure}

An undirected network of country collaboration in Fig.\ref{fig6} (a) shows that how different countries have come together to explore the potential of blockchain in different research areas as well as technologies.  The network consists of 106 nodes and 4009 edges. The nodes in the network represent the countries and the edges represent the possible collaborations between two countries.  A link appears between two nodes when the two countries have a joint publication. The size of the node reflects the number of connections a country has with others (degree) and so as the label of the node.  The width of the edge reflects the number of times the same two countries collaborate (strength).
Among all,  GBR collaborated with a large number of countries (degree=69) followed by the USA (68), China (65), Australia (56), India (52), and so on. 10.9\% countries collaborated with a single country,  and 58.1\% countries collaborated with more than 10 countries. 

106 countries across the globe have contributed to the research on blockchain.  Fig.\ref{fig6} (b) shows the contribution of top 20 countries in terms of (i) number of publications (sorted),  (ii) the number of authors,  (iii) the proportion (in \%) of lead authors (corresponding author), and (iv \& v) the proportion (in \%) of authors lead the national collaboration and international collaboration.  After giving equal credit to multi-affiliated publications, the total number of publications become 9550. The total number of lead authors is 7709 and 23.9\% of lead authors are only from China.
Publication contribution of top countries includes China (18.2\%), USA (13\%), GBR (5.8\%), India (5.4\%), South Korea (4\%), Australia (3.8\%), Canada (3.4\%), Germany and Italy (3.3\%),  Spain and Russia (2.3\%), and others <2\%.
The proportion of authors and lead authors (corresponding authors) of top countries include China (22.8\%,  23.9\%), USA (11.6\%, 11.2\%), GBR (4.5\%,  4.2\%), India (6.4\%, 5.6\%), South Korea (3.9\%, 4.9\%),  Australia (2.8\%, 3.6\%), and so on.

In Fig.\ref{fig6} (b), we also looked into the the distribution of authors within the country.  Out of 4016 authors from China, 45.8\% authors acted as lead authors (both nationally and internationally) and 35.1\% has lead the international collaboration. Similarly, in USA out of 2066 authors, 41.7\% of authors acted as lead authors and 25.2\% lead the international collaboration. In case of GBR,  out of 799 authors, 45.4\% of authors acted as lead authors and 45.2\% lead the international collaboration, which is highest among to three.

\begin{table}[h!]
\caption{Network Statistics}
\begin{tabular}{|l|c|l|c|c|}
\hline
\multicolumn{5}{|c|}{\textbf{Network Statistics}}                                                                 \\ \hline
Nodes                       & 106  &  & \multicolumn{1}{l|}{No. of Clusters}                & 5                   \\ \cline{1-2} \cline{4-5} 
Edges                       & 4009 &  & \multicolumn{2}{c|}{\textbf{Country-wise Lead Authors {[}Top 5 in \%{]}}} \\ \cline{1-2} \cline{4-5} 
Avg. Degree                 & 17.8 &  & China                                               & 21.6                \\ \cline{1-2} \cline{4-5} 
Avg. Weighted Degree        & 75.8 &  & USA                                                 & 11.9                \\ \cline{1-2} \cline{4-5} 
Network Diameter            & 5    &  & India                                               & 5.9                 \\ \cline{1-2} \cline{4-5} 
Avg. Path Length            & 2.1  &  & South Korea                                         & 5.2                 \\ \cline{1-2} \cline{4-5} 
Avg. Clustering Coefficient & 0.68 &  & UK                                                  & 4.5                 \\ \cline{1-2} \cline{4-5} 
\end{tabular}
\label{table3}
\end{table}

\subsection{Ranking based on national and international collaboration}
\begin{figure}[h!]
\centerline{\includegraphics[width=0.95\linewidth]{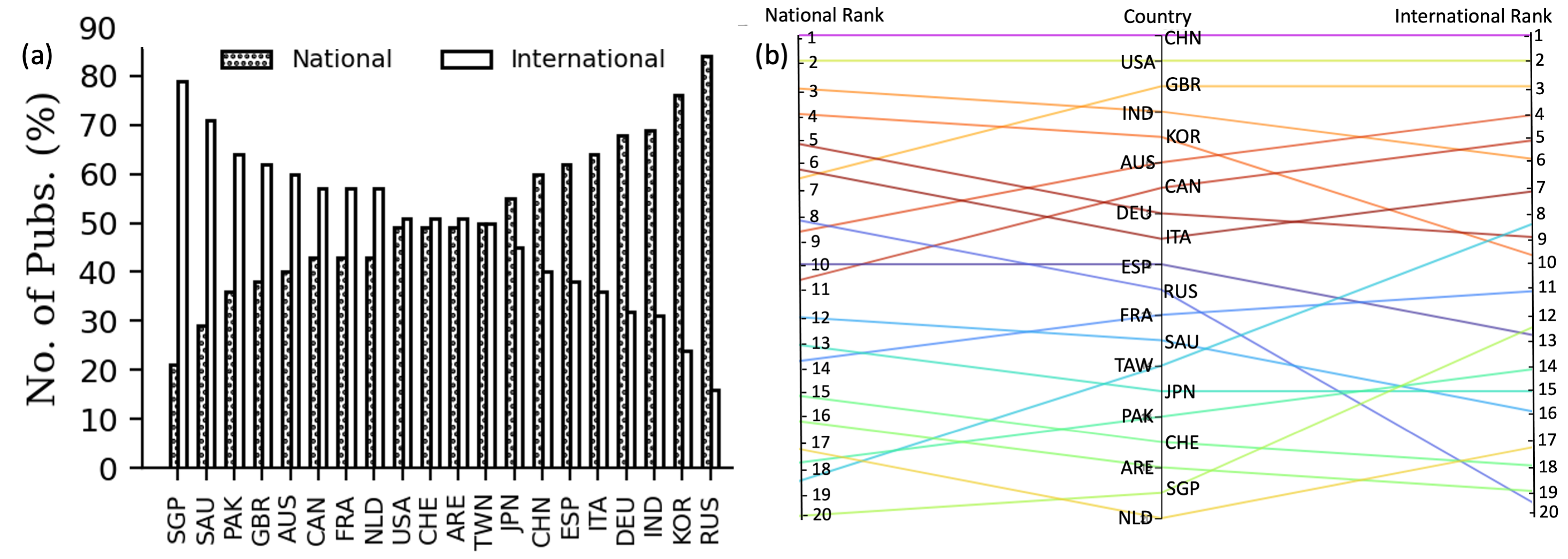}}
\caption{(a) Bar plot represents the country-wise proportion of international and national collaboration [in\%] (top 20 in number of publications). The data is sorted based on the international contribution.  (b) Ranking of top 20 countries based on national and international collaboration. }
\label{fig7}
\end{figure}

Porter et al.  \citep{porter2007measuring} proposed a way to measure research interdisciplinarity based on the relation between the subject categories.
Here, we analyzed the national and international collaboration of the top 20 countries based on the number of publications produced by all individual countries.  We calculated the national and international ranking of countries based on the share of national and international collaborations paired in the total publications of the individual country.
China, India, South Korea, Germany, Italy, Spain, Russia, and Japan have collaborated more nationally whereas GBR, Australia, Canada, France,  Saudi Arabia, Pakistan, and Singapore share more international collaboration based on collaboration share in the number of publications produced by them (see Fig.~\ref{fig7}). The plot is sorted in ascending order based on the number of international collaborations.

Singapore published a total of 121 publications and out of this, the share of international collaboration is 83.6\%, i.e., Singapore shared more international collaboration out of its total publications. The second highest is Saudi Arabia with 75.9\% of international collaboration out of a total of 160 publications. The third highest is GBR with 69.5\% of international collaboration out of a total of 562 publications. On the other hand, the USA retained the balance between national and international collaboration. In terms of national collaboration, Russia published 222 publications with a national share of 79.3\%. The second highest is South Korea with 71.4\% of national collaboration out of a total of 390 publications, and the third-highest in India with 63.9\% of national collaboration out of a total of 515 publications.

Fig.\ref{fig7} (b) shows the ranking of the top 20 countries based on national and international collaboration.  The rank is calculated based on the share of national and international collaboration out of the total number of publications by 20 countries.  China is higher in the number of publications, authors, and lead authors, so it ranked $1^{st}$ both nationally and internationally.  While maintaining the balance between national and international collaboration,  USA ranked $2^{nd}$ both nationally and internationally.  GBR ranked $7^{th}$ in national and $3^{rd}$ in international collaboration.  India secured ranked $3^{rd}$ nationally and $6^{th}$ internationally. On the other hand,  Russia ranked last in international and SGP ranked last in national collaboration.

\subsection{Thematic analysis: Mapping of countries and  keywords}
\begin{figure}[h!]
\centerline{\includegraphics[width=0.95\linewidth]{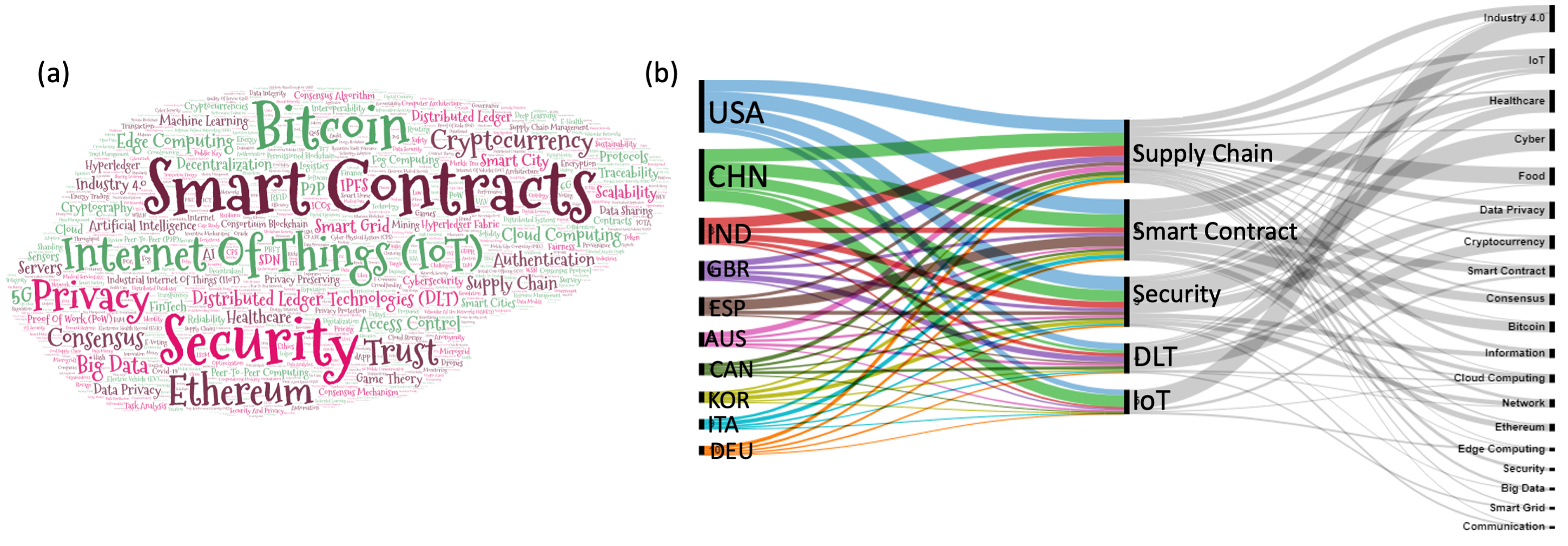}}
\caption{(a) Word cloud of author's keywords. The size of the keyword is proportional to the frequent use of the keyword.  (b) The alluvial plot shows the country-wise thematic analysis (top 10 in the number of publications).  The association of five major keywords with other keywords is shown. }
\label{fig8}
\end{figure}

Identifying the main features of the theme~\citep{velez2020thematical} and the flow of the themes within applications can be understood well through thematic analysis~\citep{braun2006using, vaismoradi2013content}. The thematic analysis identifies, analyzes,  and reports the theme (patterns) within the data~\citep{chen2008thematic}.
Author keywords are the common units of analysis, reflectors, and research evidence of the studies. The broader sense of the research topic and the area of study can be identified through the author's keywords.
Fig.\ref{fig8} shows the thematic analysis of such keywords in the form of the word cloud and alluvial diagram.
After excluding the keyword ``Blockchain'' which is our main keyword of search, a word cloud of the rest of the author's keywords is shown in Fig.\ref{fig8} (a). The size of each word reflects the frequent use (count) of that particular keyword. The highlighted keywords in the study except blockchain are \textit{Smart Contracts, Bitcoin, Internet of Things (IoT), Security,  and Distributed Ledger Technology (DLT).}

Further, the keywords used in the top 10 countries with the highest number of publications are extracted. We have identified that keywords like supply chain, smart contract, security, DLT, and IoT are used frequently with other keywords like industry 4.0, healthcare, cyber, data privacy, cryptocurrency, consensus, etc.
Hence,  a relationship between the top 10 countries and five major themes along with the further level of keywords usage is shown in Fig.\ref{fig8} (b). The choice of the top 10 countries here is just to show the implementation of five major themes along with other keywords in different countries and these countries have a large number of publications as mentioned in \ref{country}.
Also, an interesting fact of blockchain publication is that the number of authors have explored the potential of this technology with recent trends like AI, IoT, 5G, and cyber security, etc.~\citep{ante2020smart, fosso2020bitcoin}.

\section{Discussion and conclusion}
\label{sec5}
The present study presents the bibliometrics analysis on the blockchain that covers many gaps in the literature. This paper provides insight and meaningful implications regarding the popularity of blockchain across the globe from 2014-2020 with different perspectives (i) impact of open access publications on the growth and popularity of blockchain; (ii) to investigate the patterns of author's contribution, collaboration strength, and team size; (iii) to identify the ranking of countries based on the national and international collaboration; and (iv) to perform thematic analysis on the literature to investigate the major themes. 

As blockchains become more mature, this study also introduces its real-world scenario in a comprehensive way. Most of the reviewed publications agree on the promising advantages of introducing blockchain in different sectors~\citep{ante2020place}. Combining blockchain with current application trends can certainly improve the efficiency of existing systems~\citep{yu2020knowledge}.  By identifying and analyzing the most related papers, the key findings of the work can be summarized as follows:

\begin{enumerate}
\item The bibliometric analysis on 6790 publications on blockchain from 2014-2020 is performed.  In the temporal distribution of the publications, the continuous growth has been observed since 2014.  The trend shoes the rise of 4.26\% in 2016 and gone up five times in 2017.  In 2020,  a minor decay of 0.5\% is being observed in the publication count. In 2016,  25\% publications were open accessed in nature and its impact can be observed in the number of citations later in 2020. The average citations received by OA publications is 1.5 times more than the NOA. During 2017-18, OA count has decreased but during 2019-20 it has gained the momentum. In 2019, the share of OA publications observed is 26.8\% and 40.7\% in 2020. Base on this,  in 2016 the average number of citations received by NOA publications is 54.7 and OA is 45.3.  As we can see that 25\% of OA publications has on average 45.3 citations which itself reveals the impact of OA.  Similarly in 2019, 26.8\% of OA publications received on average 41.6 citations.

\item Papers on blockchain were published in 17 languages worldwide; however, the major publications are in the English language followed by Spanish and Russian.  The broader publication categories are (i) articles (49.7\%),  (ii) conference proceedings (44\%), and (iii) review (4.4\%). The average number of citations received by articles is 9.1, proceedings as 3.9 and reviews as 16.9.  Although the contribution as proceedings is nearly equivalent to articles, however, the average citations of proceeding articles is very low as compared to articles. On the other hand, the review has the lowest count but the highest average citations among the three. 

\item The paper's influence is measure through the usage count and citations provided by WoS.  It provides insight into how much interest a publication has generated among users of the platform. The results shows that the visibility of the publication is twice of the actual citations, i.e. the average number of citations received by blockchain-based publications is 7.1 and the average number of downloads received by blockchain-based publications is 16.3. Overall only 2.9\% of publications are highly cited that received more than 50 citations which are mainly as articles and reviews.  Mainly, the highly cited papers have been viewed 2.5 times than received citations.

\item The contribution of total of 14748 authors has been analyzed in author's collaboration network.  The analysis reflects that 73.7\% authors have published a single paper on blockchain till now, and 14\% published two papers whereas more than 10 papers have been published by $<1\%$ of authors.  The analysis of collaboration strength reflects that 73.5\% of author pair are such that have published a single paper, 14\% of author pair has published two papers,  and 0.5\% has published more than 10 papers together. This shows that a long and strong collaboration in the filed is almost negligible. Further the patterns of team size highlights that the 38.7\% of publications are written by single author, and 28.9\% are written by a team of two authors.  Only 3.5\% of publications are produced by the team of more than 5 authors. This show that people prefer to work in small groups.

\item Further the contribution of 106 countries in the field was analyzed through country collaboration network.  The results highlight that 10.9\% of countries have collaborated with a single country, and 58.1\% countries have collaborated with more than 10 countries.  Among all, GBR collaborated with a large number of countries followed by the USA, China, Australia, and India.  Overall, Chain is leading in the number of publications followed by the USA (13\%), GBR, India, and so on. Similarly, the number of authors are also lead by China (22.8\%) followed by the USA  India,  GBR, and so on.

\item The national and international collaboration of countries is analyzed by the number of publications produced by every individual country.  China is higher in the number of publications, authors, and lead authors, so it ranked $1^{st}$ both nationally and internationally.  While maintaining the balance between national and international collaboration,  USA ranked $2^{nd}$ both nationally and internationally.  GBR ranked $7^{th}$ in national and $3^{rd}$ in international collaboration.  India secured ranked $3^{rd}$ in nationally and $6^{th}$ internationally. On the other hand,  Russia ranked last in international and SGP ranked last in national collaboration.

\item At last, a thematic analysis is performed on the literature to investigate the major themes.  A thematic analysis observed  five major themes as supply chain, smart contract, security, DLT, and IoT. we further observed that the five major themes are used quite often with other keywords like industry 4.0, healthcare, cyber, data privacy, cryptocurrency, consensus, etc. Hence., the mapping of these observed themes has been analyzed in top 10 countries. An interesting fact of blockchain research is the fact that a number of authors have explored the potential of this technology with recent trends like AI, IoT, 5G, and cyber security, etc.

\end{enumerate}
\subsection{Limitations and future work}
The limitations and further expansion of the above study can be discussed as:
\begin{itemize}
\item  The choice of the dataset~\citep{martin2019google}: We have collected the data from WoS for the study whereas the number of records varies among the different datasets, hence the study can be extended with the data from Google Scholar, Scopus, Dimension, etc.

\item Author collaborations: The present study analyzes the author collaboration with the help of network analysis but it does not explore the gender disparity to analyze the contribution and collaboration gender-wise. Hence, the study can be extended gender-wise to analyze the role of women in the development of blockchain technology.

\item Application development: The study analyzed the literature based on types of publications but lacks the in-depth analysis of the ideas introduced by the authors. For example, whether the publication is on applications of blockchain or it’s on consensus algorithms etc. Hence, the study can be extended based on the kind of idea an author has presented and citations received by the idea.

\end{itemize}

\textbf{Authors contribution:} Both the authors conceived and designed the analysis, collected the data, performed the analysis and wrote the draft.
 \section*{Conflict of interest}
 The author declare that they have no conflict of interest.
\bibliographystyle{spbasic}      
\bibliography{cas-refs.bib}   

\end{document}